\documentclass{aa}
\usepackage{natbib}
\usepackage{epsfig}
\usepackage{csquotes}
\usepackage{tikz}
\usetikzlibrary{decorations.pathmorphing}
\usetikzlibrary{decorations.pathmorphing,calc}
\def\ut#1{\mathop{\vtop{\ialign{##\crcr
     $\hfil\displaystyle{#1}\hfil$\crcr\noalign
     {\kern1pt\nointerlineskip}\hbox{$\hfil\sim\hfil$}\crcr
     \noalign{\kern1pt}}}}}

\def\undersymbol#1#2{\mathop{\vtop{\ialign{##\crcr
     $\hfil\displaystyle{#2}\hfil$\crcr\noalign
     {\kern1pt\nointerlineskip}\hbox{$\hfil#1\hfil$}\crcr
     \noalign{\kern1pt}}}}}

\usepackage{graphicx}
\begin{document}

\title{The stability of voids in the Local Universe: The role of the cosmological constant }
       \author{V.G.Gurzadyan\inst{1,2}, N.N.Fimin\inst{3}, V.M.Chechetkin\inst{3} }

              \institute{Center for Cosmology and Astrophysics, Alikhanian National
Laboratory and Yerevan State University, 0025 Yerevan, Armenia \and
SIA, Sapienza University of Rome, 00191 Rome, Italy \and Keldysh Institute of Applied Mathematics of RAS, 125047 Moscow, Russia}

   \offprints{V.G. Gurzadyan, \email{gurzadyan@yerphi.am}}
   \date{Submitted: XXX; Accepted: XXX}

 \abstract{ The Vlasov kinetic formalism is employed to study the evolution and stability of cosmic voids in the Local Universe, taking into account not only the gravitational attraction, but also the repulsive effect of the cosmological constant (i.e., local dark energy). In accordance with the theorem on the general function of the identity between the gravitational fields of a sphere and a point mass, the cosmological constant provides a natural explanation for the Hubble tension, attributing it to local and global flows characterized by different Hubble parameters. The crucial role of the $\Lambda$-repulsion in maintaining the stability of voids at the present epoch is demonstrated when Landau damping suppresses discrete collapse modes and prevents random local density perturbations inside the voids from growing and incorporating new galaxies into the walls. Inside the voids, the $\Lambda$-repulsion exceeds the attractive force of the residual matter, driving matter outward and accelerating its migration toward the void boundaries. In the Local (late) Universe, cosmic voids have entered a stage characterized by stable and more pronounced walls, as studied via observational surveys across different redshift ranges.
}


   \authorrunning{V.G. Gurzadyan, N.N.Fimin, V.M.Chechetkin}
   \titlerunning{The stability of voids in the Local Universe}
   \maketitle
%
%________________________________________________________________

\section{Introduction}

Differences in the characterization of the early and late Universe remain among the key targets of theoretical studies and observational surveys. The Hubble tension \citep{R1,R2,R3,LR,Mon,Val,Dai,Cas,Ban1,Jia} has further highlighted this principal issue, attracting a vast number of theoretical approaches and models. Another important issue is the possible variations in dark energy with redshift  \citep{DESI1,DESI2,DES,Cap,Pop,Ong,Ban2}, which requires both more thorough statistical analyses and additional observational surveys, particularly at high redshifts. 

The formation of matter structures in the early and late Universe may also exhibit fundamental differences, reflecting the dominance of different physical processes at different scales. On cosmological scales, the Zeldovich pancake theory \citep{Z,Arn,SZ} describes the evolution of primordial density fluctuations and predicts the emergence of the large-scale cosmic web, including quasi-one-dimensional filaments. At these scales, the Zeldovich approach neglects the mutual gravitational interactions between filaments.
 
On local scales, gravitational interactions within matter can play a crucial role in subsequent structure formation and evolution. In this study, we first examine the role of self-consistent gravitational interactions in the Local Universe by means of the Vlasov kinetic approach, extending our previous studies \citep{GFC1,GFC2,GFC3,GFC4,GFC5}. Next, we shift to the central aspect of our approach, which is the notion that structure formation is governed not only by mutual gravitational attraction, but also by the repulsive effect associated with the cosmological constant. Within the non-relativistic framework, the appearance of the cosmological constant on local scales follows from the theorem \citep{G1}; see also \citep{Car} for a discussion of the theorem and pedagogical derivation. This theorem states that the most general force law preserving the equivalence between the gravitational field of a sphere and that of a point mass takes the form
\begin{equation}
F=-{\frac{\gamma Mm}{r^{2}}}+{\frac{\Lambda c^{2}mr}{3}}.
\end{equation}
The constant $\Lambda$ on the left-hand side corresponds to the cosmological constant in weak-field general relativity and in the non-relativistic cosmological model of McCrea and Milne \citep{MM}. Zeldovich in \citep{Z81} outlined the efficiency of the non-relativistic description of the Local Universe. 

Notably, Eq.~(1) predicts a non-zero force field inside a shell besides its center, whereas the $1/r^2$ law predicts a force-free field within a shell. Equation~(1) has been shown to fit observational data for pairs, groups, and clusters of galaxies \citep{GS2,G2,GS3,GS4}. The Hubble tension has also been argued to arise naturally from the distinction between local and global galaxy flows, namely, described by Eq.~(1) on local scales and by the Friedmann equations on global scales, leading to non-coinciding values of the Hubble parameter. Moreover, this framework allows one to derive absolute lower and upper bounds on the local Hubble parameter that are also consistent with observations \citep{GS7,GS8}; see also \citep{RG}. Regarding the non-force-free field within a shell following from Eq. (1),  we mention the observational indications of the influence of a spherical halo on the disk structure of spiral galaxies \citep{Kr}. It has been shown that Bose-Einstein condensate (BEC)-based technologies enable tests of the law in Eq.~(1) at the tabletop scale \citep{PRR}.

The kinetic approach and the methods developed in plasma physics were widely applied to the study of gravitating systems, while accounting for the differences between plasmas and gravitational media arising from the long-range nature of gravitational interactions \citep{Lynden1,ZN,Frid,vankampen3,Lynden2,Vand}. Vlasov’s kinetic formalism \citep{V1,V2} was also extensively employed in studies of gravitational dynamics and cosmological problems (e.g., \citealt{Maslov,Maslov1,Rubin,Coles,Dolb,Muk,vneshnie2,vneshnie1,Do,vneshnie4}).
 
In this paper, we consider the formation of large-scale structures by analysing the non-dissipative solutions of the Vlasov-Poisson equations of the van Kampen wave type \citep{VK}. We show that the periodicity of these waves is violated when the repulsive force associated with the cosmological term is taken into account, since that would lead the system to get close (in local terms) to a weakly inhomogeneous one. The repulsive effect of the cosmological constant (or dark energy on local scales) appears to play a crucial role in suppressing gravitational instability, preventing perturbations on those scales from growing, and thereby implying that cosmic voids do not collapse, but that they actually remain stable.

\section{The cosmological term and the linearization of the Vlasov-Poisson system}

To describe the evolutionary dynamics of cosmological structures in the Local Universe we consider
the self-consistent gravitational field approximation for a set of $N$ ``particles'' with masses of $m_{i=1,...,N} =m\equiv 1$, which could be galaxies, galaxy groups, or even clusters. The closed (nonlinear) system of Vlasov--Poisson equations taking into account the anti-gravitational influence of dark energy, by
including the cosmological term $\Lambda$ in the consideration; in this case, can be represented \citep{GFC5} as
\begin{equation}
\frac{\partial F({\bf x},{\bf v},t)}{\partial t} +
{\nabla}_{\bf x}({\bf v}F)+\widehat{\mathcal{G}}_\Lambda(F;F)=0,
\end{equation}
$$
\widehat{\mathcal{G}}_\Lambda(F; F) \equiv
-{\nabla}_{\bf v}F\cdot {\nabla }_{\bf x}\Phi [F({\bf x},{\bf v},t)]),
\label{1}
$$
\begin{equation}
{\Delta}_{\bf x}\Phi[F({\bf x})]=
4\pi A \gamma \int F( {\bf x},{\bf v},t)\:d{\bf v}-c^2\Lambda,
\label{2}
\end{equation}
\noindent
where $F({\bf x},{\bf v},t)$ is the distribution function of gravitational interacting particles,
$A$ is a normalization factor for a particle density of $A=1,n_0,N$ depending on the choice of the number of gravitationally interacting particles), $\gamma$ is the gravitational constant. The system of particles is situated in large domain of configuration space, $\Omega \subset {\mathbb R}^3_{\bf x}$ (${{\rm{diam}}\:\Omega \equiv R_\Omega}< \infty$). At the boundary of the $\partial \Omega$ region, the value of the total potential is considered limited.

Equation (\ref{2}) is a nonuniform Poisson equation for a self-consistent generalized Newton--type gravitational potential $\Phi[F({\bf x})]$. In the case of an a priori known dependence of the potential on the particle density
distribution (e.g., of the Maxwell-Boltzmann type), Eq. (\ref{2}) becomes closed. After linearization in the neighborhood of the chosen stationary solution of the Vlasov equation, the value of $\Phi({\bf x})$ can be determined by solving the Dirichlet boundary value problem with given conditions on the inner surface $\partial \Omega$.
It is assumed that in this case the condition of the theorem in \citep{G1} is satisfied; this approach was previously considered in \citep{GFC4,GFC5}). We assume that the linearization of the Vlasov-Poisson system of equations can be performed separately for the kinetic equation expressed via Eq. (\ref{1}) and for the Poisson equation expressed via Eq. (\ref{2}).
The Maxwellian distribution $f_0({\bf v})=(2\pi T)^{-3/2}\exp\big(-{\bf v}^2/(2T)\big)$ ($F\propto n_0 f_0 +o(f_0)$) is usually chosen as the stationary solution or a multi-Maxwellian in the general case, with its functions depending only on the velocity variables.

The gravitational field potential taking into account the symmetries of the formulation of additional conditions and allowing the term to be replaced with the Green's function of the boundary value problem with the constant $C_1$ yields
\begin{equation}
\bigg(\widehat{I}-\kappa \widehat{\mathfrak{G}}\bigg)\Phi =-\frac{c^2\Lambda}{6}|{\bf x}|^2 +{\mathfrak{T}}_0,
\end{equation}
$$
\widehat{\mathfrak{G}}(\Phi) \equiv -\int \frac{\exp\big( -{\Phi ({\bf x}')}/{T} \big)}{|{\bf x}-{\bf x}'|} \:d{\bf x}',
\label{3}
$$
$$
\kappa \equiv A^\dag \gamma,~~~
A^\dag = 4\pi A\int_0^{v_{max}} \exp\big(-\frac{y^2}{2T}\big)y^2dy,
$$
$$
{\mathfrak{T}}_0 = -\frac{\gamma n_0}{R_\Omega}-\frac{c^2\Lambda R_\Omega^2}{6}+C_1.
$$
In general case, we obtained an inhomogeneous Hammerstein integral equation with respect to the
variable $\Phi=\Phi([F]\big|_{F =F ({\bf x},{\bf v})}$). The homogeneous equation has a solution in the form of a constant potential $\Phi({\bf x})\to C$ which obviously extends beyond the boundary of the domain.
While the inhomogeneous equation has a locally unique solution, since the integral weakly polar
operator is compact and the free term is bounded in the domain $\Omega$. The solution is extended in such a way that the right-hand side in (\ref{3}) is bounded outside the domain and the a priori requirement of smoothness of the potential must be satisfied. Considering the homogeneous Fredholm equation of the second kind associated with Eq. (\ref{3}) with a linearized integral operator with respect to the perturbation $\phi=\Phi - C$, when the term with the cosmological constant cannot be positioned as a perturbation, expressed as
$\big(\widehat{I}-\kappa \widehat{\mathfrak{G}}'\big)\phi({\bf x})=0$, whose solution takes the form
\begin{equation}
\phi({\bf x})=-\kappa^\dag\int \frac{\phi ({\bf x}')}{|{\bf x}-{\bf x}'|} \:d{\bf x}', ~~~\kappa^\dag\equiv \kappa \exp(-C).
\label{4}
\end{equation}
It should be specifically emphasized that in this case, we actually obtained a problem on eigenvalues and eigenfunctions for an unbounded space, and not for a bounded domain $\Omega$. This is because we artificially introduced a perturbation of the solution to the nonlinear equation without imposing additional constraints on it at the boundary. In plasma physics, assumptions of this kind are quite common, which have led to the use of this method also for problems involving gravitational interactions. This can be justified for a decreasing (in absolute value) potential; however, for an increasing (again in absolute value) potential (due to the presence of an additional constant term $c^2\Lambda$ on the right-hand side of the Poisson equation), the method requires the use of a de-singularizing transformation
of independent phase space coordinates or the abandonment of the locality of the normal mode values.
Both approaches have advantages and disadvantages, but the authors believe that the lens transformation method can lead to a plethora of physically unrealizable situations that are formally admissible but violate ``Occam's razor;'' this analogy can be drawn with the Kruskal-Szekeres ``maximal extension'' of the Schwarzschild metric. Therefore, we chose to focus on studying the properties of the  ``standard''  linearized Vlasov equation in the representation of coordinates of the extended phase space $({\bf x},{\bf v},t)$, which is Fourier-transformed and Laplace-transformed, allowing us to determine solutions of this equation in terms of modified non-local normal modes.

Representing the solution of system (\ref{1})--(\ref{2}) in the form $F({\bf x},{\bf v},t)=F_0({\bf x},{\bf v})+f({\bf x},{\bf v},t)$,
$\Phi [F]=\Phi_0[F_0] +\phi[f({\bf x},{\bf v},t)]$ at the condition $|F_0 |\gg | f |$. Here, $F_0$ is a stationary-equilibrium solution of the Vlasov equation, which, generally speaking, in the presence of an external force is not a Maxwellian distribution. Excluding any terms on an order that is higher than linear in $f$, we obtained

\begin{equation}
\frac{\partial f}{\partial t} + {\bf v}\nabla_{{\bf x}}{f}- \nabla_{{\bf x}}\phi({\bf x}) \cdot\nabla_{{\bf v}}{F_0}+
\big(-\nabla\Phi_0 [F_0] + \frac{1}{3}c^2\Lambda {\bf x} \big)\cdot \frac{\partial f}{\partial {\bf v}}=0.
\label{5}
\end{equation}
\noindent
Obviously, in this case, $\Phi_0[F_0]={\rm const}$ only for $F_0=F_0({\bf v})$ due to the spatial homogeneity of the particle distribution far from the boundaries of the $\Omega$ region. Then, if we take into account the degeneracy (into a constant value) of the Green's function for the boundary value problem for the gravitational potential, then
the fourth term in the linearized Vlasov equation takes the form $({\bf G}_{sc}+c^2\Lambda {\bf x}/3)\cdot f_{\bf v}'$ (${\bf G}_{sc}$ is the strength of the self-consistent gravitational field in a multi-particle system.

Thus, the form of the linearized equation describing the self-consistent particle-attraction field, including the influence of the cosmological term, differs substantially from its counterpart in plasma theory for charged particles. This necessitates a detailed investigation of gravitational analogs of van Kampen waves and Landau damping \citep{vankampen3}, as well as a reformulation of fundamental concepts such as the permittivity function and the dispersion relation.

In what follows, we analyzed the linearized Vlasov-Poisson system modified by the cosmological term in an infinite spatial domain. Our objective is to derive the corresponding dispersion relations and examine their physical implications, including the conditions for the occurrence of Landau damping and Jeans instability. These processes may play an important role in shaping the observed matter distribution and structure formation in the Local Universe.

\section{Equilibrium configurations}

In this section, we investigated the system of Eqs. (\ref{1}) and (\ref{2}) governing perturbations of the equilibrium distribution function. We assumed that the homogeneous Poisson equation for the perturbation of the gravitational potential; namely, the self-consistent attractive field, depends linearly on the density perturbation. This assumption enables us to treat the repulsive-force term in the kinetic equation explicitly and to directly apply the superposition principle to gravitational fields arising from different sources.

For the perturbations of the distribution function, we adopted a linearized kinetic equation in the form (\ref{5}), and for the perturbation of the potential in the form $\nabla^2\phi({\bf r};t)=4\pi\gamma\int f\:d{\bf v}$. If the equilibrium state is assumed to be spatially homogeneous, namely, in the Jeans approximation, where $\nabla_{\mathbf{x}} F_0 = 0$, and the background forces yield $\nabla_{\mathbf{x}} \Phi_0 = c^2\Lambda {\bf x}/3$. In other words, they cancel each other out or are not taken into account on a local scale, the equation is simplified to a comprise a characteristic of plasma physics: $\partial f/\partial t+{\bf v}\nabla_{\bf x}f-
\nabla_{\bf x}\phi\cdot\nabla_{\bf v}F_0=0$.

It should be noted that the general form of the unperturbed state of the particle background is given by the function $F_0^{(gen)}({\bf x},{\bf v})$, which is a stationary solution ($\partial/\partial t \equiv 0$) of the Vlasov equation. Therefore, the Maxwellian $F_0({\bf v})$ is an important (albeit special case), which is used below to avoid cumbersome calculations and to demonstrate an important ``limiting case,'' corresponding to the extremum point of the total gravitational potential.
Its use can also be justified under the conditions of applicability of the so-called ``Jeans swindle.'' Specifically, at the quasi-homogeneity of the medium on small local scales for short wavelengths when it is formally assumed that at zeroth order, the gravitational force of the background and the external force completely compensate for each other. In the general case, we have
\begin{equation}
F_0({\bf x},{\bf v})\:\to\:
F_0^{(gen)}({\bf x},{\bf v})=
\end{equation}
$$
n_c \exp\bigg( -\frac{m }{2\pi k_B T} \bigg)^{3/2}
\exp\bigg( -\frac{m \Phi_0({\bf x})}{k_B T} + \frac{m c^2 \Lambda {\bf x}^2}{6 k_B T} - \frac{m {\bf v}^2}{2 k_B T} \bigg),~~~
\label{6}
$$
\noindent
where $n_c$ is the density at the point where $U_{eff}({\bf x})\equiv\Phi_0({\bf x})-{c^2\Lambda}|{\bf x}|^2/{6} =0$; obviously, the magnitude of the effective potential satisfies Eq. (\ref{2}). We note that the system of particles to be considered in a quasi-stationary state to avoid the escape of particles to infinity under the action of a repulsive potential of unlimited magnitude or the system should be considered in a bounded domain, $\Omega$.

We considered the derivation of a linear Fredholm integral equation of the third kind from the linearized Vlasov-Poisson system. We represented the perturbation of the distribution function \(f(\mathbf{x}, \mathbf{v}, t)\) and the fluctuation of the gravitational potential,\(\phi(\mathbf{x}, t)\), as normal modes. In the perturbation, \(f\), we isolated the spatio-temporal exponential and certain (as yet unknown) structure function of the velocities \(\Psi(\mathbf{v})\): \(f(\mathbf{x},\mathbf{v},t)=\Psi (\mathbf{v})\cdot \exp (i\mathbf{k}\cdot \mathbf{x}-i\omega t)\), \(\phi (\mathbf{x},t)=
\phi_{\omega,\mathbf{k}}\cdot \exp(i\mathbf{k}\cdot \mathbf{x}-i\omega t)\). We substituted these expressions into the linearized Vlasov equation. Differentiating the exponentials with respect to time and coordinates, in the local (quasi-homogeneous) WKB approximation, we obtained a rigorous algebraic relation for the functions when the locally homogeneous approximation is valid for short waves and the scale of the perturbation is much smaller than the scale of the background inhomogeneity $kL\gg 1$ (see Sect. 4). Then, we have
\begin{equation}
(\mathbf{k}\cdot \mathbf{v}-\omega )\Psi (\mathbf{v})=-(\mathbf{k}\cdot
 \mathbf{v})\frac{m\phi_{\omega,\mathbf{k}}}{k_{B}T}F_{0}(\mathbf{x},\mathbf{v}).
\label{7}
\end{equation}
\noindent
Next, we want to direct the coordinate axis along the wave vector \(\mathbf{k}\). Let \(u = {\mathbf{k}\cdot\mathbf{v}}/{k}\) be the projection of the particle velocity and \(v_{ph} = \omega/k\) be the phase velocity of the mode. We integrate the function \(\Psi(\mathbf{v})\) over the transverse velocities,\(\mathbf{v}_{\perp}\), to obtain the 1D profile of our normal mode, which we denote as
\(\psi(u) = \int \Psi(\mathbf{v}) d^2\mathbf{v}_{\perp}\). Integrating the Vlasov equation over the transverse velocities,
we find \((ku-kv_{ph})\psi (u)=-ku {m\phi_{\omega,\mathbf{k}}}/({k_{B}T})\cdot F_{0}(\mathbf{x},u)\). Divide by wave number \(k\): \((u-v_{ph})\psi (u)=-{m \phi_{\omega,\mathbf{k}}}/({k_{B}T})\cdot u\cdot
F_{0}(\mathbf{x},u)\), where \(F_0(\mathbf{x}, u)\) is a 1D Maxwell-Boltzmann distribution preserving the term
extensions,
$$
F_0(\mathbf{x}, u) = n_c \left( \frac{m}{2\pi k_B T} \right)^{1/2},
$$
$$
 \exp\left( -\frac{m \Phi_0(\mathbf{x})}{k_B T} + \frac{m c^2
\Lambda \mathbf{x}^2}{6 k_B T} - \frac{m u^2}{2k_B T}\right).
$$
The amplitude of the potential perturbation \(\phi_{\omega, \mathbf{k}}\) is not an independent constant; it is directly
related to the profile of our normal mode \(\psi(u)\) through the Poisson equation, \(-k^{2} \phi_{\omega,\mathbf{k}}=4\pi \gamma m\int_{-\infty }^{\infty }\psi (u^{\prime})du^{\prime}\). We substituted the expression obtained for \(\phi_{\omega, \mathbf{k}}\) back into the Vlasov equation. The minuses in front of the fractions cancel each other out and we obtained the final closed Fredholm integral equation of the third kind for the function, \(\psi(u)\),
\begin{equation}
(u-v_{ph})\psi (u)=\bigg(\frac{4\pi \gamma m^{2}}{k^{2}k_{B}T}\cdot u\cdot F_{0}(\mathbf{x},u)\bigg) \int_{-\infty }^{\infty }\psi (u^{\prime })du^{\prime }.
\label{8}
\end{equation}
\noindent
The desired function \(\psi(u)\) is preceded by a linear factor \(g(u) = v_{ph} - u\). Since the variable
\(u\) (the thermal velocity of the particles) ranges from \(-\infty\) to \(+\infty\), the function \(g(u)\)
is guaranteed to vanish at the point \(u = v_{ph}\) and by definition this reduces the equation to a third-kind equation. The right-hand side of the last relation is a Fredholm integral operator with a degenerate kernel \(K(u, u') = A(\mathbf{x}, u) \cdot 1\), where
$$A(\mathbf{x},u)=\frac{4\pi \gamma m^{2}n_{c}}{k^{2}k_{B}T}\left(\frac{m}{2\pi k_{B}T}\right)^{1/2}\cdot u\cdot 
$$
$$
\exp
\left(\underbrace{-\frac{m\Phi_{0}(\mathbf{x})}{k_{B}T}+\frac{mc^{2}\Lambda \mathbf{x}^{2}}{6k_{B}T}}_{\text{Spatial\ modulation\ background}}-\frac{m u^{2}}{2k_{B}T}\right).
$$
At the center of the system with dominating self-consistent gravity, \(\Phi_{0}\), the exponential factor reduces
the amplitude of the core, decreasing the effective intensity of the integral interaction.
At the periphery of the system where external repulsion \(c^2\Lambda {\bf x}^2\) dominates, due to the positive sign of the term
\({mc^{2}\Lambda \mathbf{x}^{2}}/{(6k_{B}T)}\), the amplitude of the core \(A(\mathbf{x}, u)\) locally increases. This means that the intensity
of the resonant interaction between the wave and the runaway particles becomes stronger with distance from the center of the gravitational condensation. The singular turning point \(u = v_{ph}\) now not only determines the attenuation, it is spatially modulated: at different points \(\mathbf{x}\), different numbers of particles enter into resonance with the same phase velocity of the wave.

We can write our Fredholm equation of the third kind for a fixed spatial point, \(\mathbf{x}\),  so the coordinate (in the framework of a local WKB analysis) acts as a fixed parameter of the medium\((v-u)\psi (u)=A(\mathbf{x},u)\int_{-\infty }^{\infty }\psi (u^{\prime })\,du^{\prime }\), where \(v \equiv v_{ph} = \omega/k\) is the phase velocity of the wave (eigenvalue of the operator) and the kernel function \(A(\mathbf{x}, u)\) explicitly contains the Maxwell-Boltzmann profile,
$$
A(\mathbf{x},u)=\frac{4\pi \gamma m^{2}}{k^{2}k_{B}T}\cdot u\cdot F_{0}(\mathbf{x},u)=\frac{\omega _{J}^{2}(\mathbf{x})}{k^{2}v_{T}^{2}}\cdot u\cdot \left(\frac{1}{2\pi v_{T}^{2}}\right)^{1/2},
$$
$$
\exp \left(-\frac{u^{2}}{2v_{T}^{2}}\right),\, v_T=\sqrt{k_B T/m}. 
$$
% (hereinafter $v_T=\sqrt{k_B T/m}$).
Here, we have 
$$
\omega_{J}^{2}(\mathbf{x})=4\pi \gamma mn_{c}\exp \left(-\frac{m\Phi_{0}(\mathbf{x})}{k_{B}T}+\frac{mc^{2}\Lambda \mathbf{x}^{2}}{6k_{B}T}\right),
$$
which is the local Jeans frequency at which the cancellation or imbalance of forces shifts the density to the periphery or to the center. We chose the standard normalization for the density perturbation, \(\int_{-\infty}^{\infty} \psi(u') \, du' = 1\), which corresponds to the van Kampen normalization, $g({\bf v})$ \citep{VK}.

Next, the Fredholm equation of the third kind can be simplified to a singular algebraic form of $(v-u) \psi_{v}(u)=A(\mathbf{x},u)$. Dividing directly by  $(v - u)$ leads to the divergence at the resonance point $u = v$, where the thermal velocity of the particle coincides with the phase velocity of the wave. The general solution of such an equation, according to van Kampen, is sought as the sum of a particular solution of the singular part; this is in the sense of the Cauchy principal value $P.V.$  and the fundamental solution of the homogeneous equation $(v - u)\psi = 0$  proportional to the Dirac delta function $\delta(v - u)$. The explicit form of a singular mode (Van Kampen wave) for a fixed eigenvalue $v$ is  $\psi_{v}(u)={P.V.}\frac{A(\mathbf{x},u)}{v-u}+\lambda (\mathbf{x},v)\delta (v-u)$,  where $\lambda(\mathbf{x}, v)$ is a function to be determined from the available additional conditions and describing the contribution of resonant particles.

To obtain the explicit form of \(\lambda(\mathbf{x}, v)\), we integrate the resulting solution \(\psi_v(u)\) over the entire velocity space and set it equal to unity: \(\int_{-\infty }^{\infty }\psi_{v}(u)\,du=
\int_{-\infty }^{\infty }{P.V.}\frac{A(\mathbf{x},u)}{v-u}\,du+\int _{-\infty }^{\infty }\lambda (\mathbf{x},v)\delta (v-u)\,du=1\). Using the filtering property of the delta-function for the second integral, we obtained \(\lambda (\mathbf{x},v)=1-{P.V.}\int_{-\infty }^{\infty }\frac{A(\mathbf{x},u)}{v-u}\,du\).
We substitute the explicit form of the Maxwell-Boltzmann kernel \(A(\mathbf{x}, u)\) with
$$
\lambda (\mathbf{x},v)=1-\frac{\omega_{J}^{2}(\mathbf{x})}{k^{2}v_{T}^{2}}{P.V.}\int_{-\infty }^{\infty }\frac{u}{v-u}\left(\frac{1}{2\pi v_{T}^{2}}\right)^{1/2},
$$
$$
\exp \left(-\frac{u^{2}}{2v_{T}^{2}}\right)\,du.
$$
This integral is expressed in terms of the real part of the plasma Kramp dispersion function \(Re \, Z(\zeta)\), where \(\zeta = {v}/({\sqrt{2}v_T})\), so that \(\lambda (\mathbf{x},v)=1-{\omega _{J}^{2}(\mathbf{x})}/{(k^{2}v_{T}^{2})}\big(1+\zeta Re\,Z(\zeta )\big)\).

Collecting all the components of the singular representation of the right-hand side of the van Kampen wave definition together, we obtained
\begin{equation}
\psi_{v}(u)=\frac{\omega_{J}^{2}(\mathbf{x})}{k^{2}v_{T}^{2}}\left(\frac{1}{2\pi v_{T}^{2}}\right)^{1/2}{P.V.}\frac{u\cdot
\exp \left(-u^{2}/2v_{T}^{2}\right)}{v-u}+
\end{equation}
$$
\left(1-\frac{\omega_{J}^{2}(\mathbf{x})}{k^{2}v_{T}^{2}}\left[1+\zeta Re\,Z(\zeta )\right]\right)\delta (v-u).
\label{9}
$$
\noindent
Here, the spatial modulation due to the cosmological repulsion, \(\Lambda \), and the background gravity, \(\Phi_0(\mathbf{x})\), is completely encapsulated in the local Jeans frequency,
$$\omega_{J}^{2}(\mathbf{x})=4\pi \gamma mn_{c}\exp \left(-{m\Phi_{0}(\mathbf{x})}/{(k_{B}T)}+{mc^{2}\Lambda \mathbf{x}^{2}}/({6k_{B}T})\right).$$

Thus, for any real value of the phase velocity \(v \in (-\infty, +\infty)\) we can uniquely determine the function \(\lambda(\mathbf{x}, v)\); the system of Vlasov--Poisson equations has a continuous spectrum of purely real modes. Van Kampen waves do not decay by themselves. Landau (or Jeans) decay of macroscopic quantities arises only upon
integration (phase mixing) of this continuous set of modes. As these modes form a complete basis
of functions, any initial perturbation \(f(t=0)\) in the system can be expanded as an integral over the singular van Kampen waves. Since the Vlasov operator is not self-adjoint in the usual Hilbert space, \(L_{2}\), the functions \(\psi_v(u)\) are not orthogonal to each other with unit weight. Orthogonality is proved using the Sokhotskii-Plemelj relations and the introduction of a dual basis. For two modes with different eigenvalues, \(v\) and \(v'\), the orthogonality condition is written in terms of the generalized weight function $1/A(\mathbf{x}, u)$ via
$$
\int_{-\infty}^{\infty }\frac{\psi_{v}(u)\psi_{v^{\prime}}(u)}{A(\mathbf{x},u)}\,du=N(\mathbf{x},v)\delta (v-v^{\prime}),
$$
$$
N(\mathbf{x}, v)=
({\lambda^{2}(\mathbf{x},v)+\pi^{2}A^{2}(\mathbf{x},v)})/{A(\mathbf{x},v)},
$$\noindent
where  the latter relation is the normalization factor of the continuous spectrum. This result shows that van Kampen modes with different phase velocities are independent (orthogonal). Inside the cosmic void, under the influence of the \(\Lambda \) term, the function \(A(\mathbf{x}, u)\) is exponentially suppressed, which leads to a sharp increase in the orthogonality weighting coefficient at the periphery of the system.

Cosmological repulsion and self-consistent gravity enter the rigorous solution (\ref{9}) via the local Jeans frequency, \(\omega_J^2(\mathbf{x})\). In regions where the cosmological term \(\Lambda \) ``repels'' matter (periphery, low density), the parameter \(\frac{\omega_J^2(\mathbf{x})}{k^2 v_T^2} \to 0\). In this limit, the first (singular) term vanishes, and \(\lambda \to 1\), transforming the mode into pure free motion of non-interacting particles (\(\psi_v(u) \to \delta(v-u)\)).
In regions of strong gravitational compression (the core of a multi-particle system, high density), the coefficient of the singular integral sharply increases, enhancing collective effects and leading to strong gravitational polarization of the medium around resonant particles.

In the Van Kampen theory, the macroscopic dispersion properties of the medium are determined by the normalization of the profile function. We used the above normalization for density fluctuations in velocity space: \(\int_{-\infty}^{\infty} \psi_v(u') \, du' = 1\). Then, the generalized singular solution (normal mode) takes the form \(\psi _{v}(u)={P.V.}\big({A({\bf x},u)/(u-v)}\big)+\lambda (\mathbf{x},v)\delta (u-v)\).
Integrating this equation over the entire velocity spectrum,
\(u\), from \(-\infty\) to \(+\infty\), we must obtain unity. This requirement for self-consistency of the Vlasov-Poisson system
generates the dispersion relation,
\(1-{P.V.}\int _{-\infty }^{\infty }{A(\mathbf{x},u)}/({u-v})\,du=\lambda (\mathbf{x},v)\).
If we are looking for discrete modes of the system, they correspond to values of the phase velocity, \(v\), at which
the contribution of free particles not associated with the field is zero. This means that in the discrete spectrum, the coefficient of the delta-function vanishes: \(\lambda(\mathbf{x}, v) = 0\).
Then the dispersion relation for discrete modes takes a strict integral form, \(\varepsilon (v,\mathbf{x})=1-{P.V.}\int_{-\infty }^{\infty }{A(\mathbf{x},u)}/({u-v})\,du=0\).

If the phase velocity of the wave is complex (\(v = v_r + i v_i\), which is equivalent to a complex frequency (and, accordingly, to Landau damping) \(\omega = \omega_r + i\gamma_L\)), the integral cannot be calculated in the Cauchy principal value sense (\({P.V.}\)) since the pole \(u = v\) departs from the real velocity axis.
For a correct description, according to Landau's bypass rule, we must analytically continue the function
\(\varepsilon(v, \mathbf{x})\) from the upper half-plane of the complex variable \(v\) to the lower half-plane.
By deforming the integration contour under the pole, we obtained the complex dispersion relation:
\(\varepsilon (v,\mathbf{x})=1-\int_{-\infty }^{\infty }{A(\mathbf{x},u)}/({u-v})\,du-2i\pi A(\mathbf{x},v)=0\). Next, we can substitute the explicit form of our Maxwell-Boltzmann kernel, \(A(\mathbf{x}, v)\), rewriting the integral part in terms of the plasma dispersion function \(Z(\zeta)\), where \(\zeta = v / (\sqrt{2}v_T)\) via
\begin{equation}
\varepsilon (v,\mathbf{x})=1-\frac{\omega_{J}^{2}(\mathbf{x})}{k^{2}v_{T}^{2}}
\big(1+\zeta Z(\zeta )\big)-2i\pi \frac{\omega_{J}^{2}(\mathbf{x})}{k^{2}v_{T}^{2}}\cdot v\cdot 
\end{equation}
$$
\left(\frac{1}{2\pi v_{T}^{2}}\right)^{1/2}
\exp\big({-\frac{v^{2}}{2v_{T}^{2}}}\big)=0.
\label{10}
$$
\noindent

The singular normal van Kampen modes \(\psi_v(u)\) with purely real \(v\) do not decay in time by themselves.
However, the macroscopic density is formed as a wave packet, that is, a superposition (integral) of an infinite number of such modes:
\(n_f \sim \int C(v)\psi_v(u) \exp({-ikvt}) dv\). Due to the phase mixing of these modes, the total field
of the packet decays. The rate of this attenuation in complex representation corresponds to the imaginary part
of the discrete root \(v = v_r + i v_i\), with the attenuation decrement \(\gamma_L = k v_i < 0\).
Assuming that the thermal velocity of the particles is small compared to the phase velocity of the wave (\(\zeta \gg 1\)),
we expanded the real part \(\varepsilon \) into an asymptotic series, and determine the imaginary
part through the Landau residue.
The condition \(\text{Re}\,\varepsilon(v_r) + i\,\text{Im}\,\varepsilon(v_r) + i v_i ({\partial \text{Re}\,\varepsilon})/{\partial v_r} = 0\)
gives an imaginary addition to the speed \(v_i = -\text{Im}\,\varepsilon / \frac{\partial \text{Re}\,\varepsilon}{\partial v_r}\).
Multiplying by \(k\), we find the Landau damping decrement:
\(\gamma_L (\mathbf{x})=-\sqrt{{\pi}/{8}}\big({\omega_{J}^{2}(\mathbf{x})\cdot
\omega_{r}^{2}}\big)/({k^{3}v_{T}^{3}})\exp\big(-\frac{\omega_{r}^{2}}{2k^{2}v_{T}^{2}}\big)\),
where the real part of the frequency satisfies the local Jeans law: \(\omega_r^2 = k^2 v_T^2 - \omega_J^2(\mathbf{x})\).

The resulting Landau damping formula for van Kampen waves demonstrates a strict spatial selection in cosmic structures with both gravity and repulsion. The repulsion \(\Lambda \) enters the damping decrement through the Jeans frequency \(\omega_J^2(\mathbf{x})\). At the periphery of systems and within voids, where the external expansion exponentially reduces the background density and the local frequency is \(\omega_J^2(\mathbf{x}) \to 0\).
Therefore, \(\gamma(\mathbf{x}) \to 0\). Collisionless damping of van Kampen waves inside voids ceases completely,
the medium becomes absolutely passive. Van Kampen waves traveling outward from the center of the gravitational concentration (\(\mathbf{x}=0\)) to the periphery enter a region where the Fredholm kernel \(A(\mathbf{x}, u)\) is modified by the \(\Lambda \) term. The background particles are accelerated by an external oscillatory potential, which continuously changes their resonance conditions with the wave and converts the Landau damping to a nonlinear dynamic regime.

To conclude our analysis of the use of Fredholm equations of the third kind, we consider the evolution
of a point initial pulse in velocity space, which requires constructing a time propagator (Green's function) for the linearized Vlasov equation. In the original physical coordinates, the local Vlasov equation for the normal mode, as shown above, is equivalent to the Fredholm integral equation of the third kind. The Cauchy problem for the evolution of an initial perturbation,
\(f(u, t=0) = \delta(u - u_0)\), where \(u_{0}\) is the initial velocity of the perturbation is solved by expanding over the complete orthonormal van Kampen basis of singular modes, \(\psi_v(u)\).
The Green's function \(G(u, u_0, t, \mathbf{x})\) satisfies the linearized Vlasov equation with a singular initial condition in the velocities, \(\frac{\partial G}{\partial t}+ikuG-ikA(\mathbf{x},u)\int_{-\infty }^{\infty }G(u^{\prime},u_{0},t,\mathbf{x})\,du^{\prime}=0\), \(G(u,u_{0},t=0,\mathbf{x})=\delta (u-u_{0})\).
Since the van Kampen modes \(\psi_v(u)\) form a complete basis for solutions of the Fredholm equation in the continuous real spectrum of eigenvalues, \(v \in (-\infty, +\infty)\), we expand the Green's function in this basis:
\(G(u,u_{0},t,\mathbf{x})=\int_{-\infty }^{\infty }C(v,u_{0})\psi _{v}(u)e^{-ikvt}\,dv\). Using the initial condition at \(t=0\), we obtained
$$\int _{-\infty }^{\infty }C(v,u_{0})\psi _{v}(u)\,dv=\delta (u-u_{0}).$$
To find the coefficients of \(C(v, u_0)\),
we applied the previously obtained van Kampen mode orthogonality condition with a
weight of \(1/A(\mathbf{x}, u)\): \(\int_{-\infty }^{\infty }{\psi_{v}(u)\psi_{v^{\prime}}(u)}/{A(\mathbf{x},u)}\,du=N(\mathbf{x},v)\delta (v-v^{\prime})\).
We multiplied both sides of the expansion by \(\psi_{v'}(u)/A(\mathbf{x}, u)\) and integrate over \(u\). Due to the orthogonality delta function, the integral on the left-hand side collapses, and we obtain the exact form of the coefficients,
$C(v,u_{0})=\frac{\psi_{v}(u_{0})}{A(\mathbf{x},u_{0})N(\mathbf{x},v)}$. Substituting the coefficient we obtained into the spectral integral led to an explicit analytical representation of the Green's function through the factors of the normal modes,
$$G(u,u_{0},t,\mathbf{x})=\int_{-\infty}^{\infty}\frac{\psi_{v}(u_{0})\psi_{v}(u)}{A(\mathbf{x},u_{0})N(\mathbf{x},v)}\exp({-ikvt})\,dv.$$

To reduce this integral to a physically perceptible form, we substituted the explicit singular structure of the modes \(\psi_v(u) = {P.V.}\big({A(\mathbf{x}, u)}/{(u - v)}\big) + \lambda(\mathbf{x}, v) \delta(u - v)\) and used the Sokhotskii--Plemelj relations.
After a series of algebraic transformations, the Green's function is separated into two fundamental terms, ballistic and collective ones, expressed as
\begin{equation}
G(u,u_{0},t,\mathbf{x})=\delta (u-u_{0})\exp\big({-iku_{0}t}\big)+,
\end{equation}
$$
\frac{A(\mathbf{x},u)}{u-u_{0}}\bigg(\frac{\exp({-ikut})}{\varepsilon^{+}(u,\mathbf{x})}-\frac{\exp({-iku_{0}t})}{\varepsilon^{-}(u_{0},\mathbf{x})}\bigg).
\label{11}
$$
\noindent
Here, \(\varepsilon^\pm(u, \mathbf{x}) = \lambda(\mathbf{x}, u) \pm i \pi A(\mathbf{x}, u)\) are the limiting values  of the complex gravitational
permittivity on the real axis, respectively, from above and below. The resulting Green's function \(G(u, u_0, t, \mathbf{x})\) describes in detail how a point perturbation of velocities evolves in coordinate space over time.

The first term \(\delta(u - u_0) \exp({-i k u_0 t})\) describes the purely kinetic behavior of particles
of the initial momentum. They move by inertia with their initial velocity, \(u_{0}\).
In space, this leads to phase mixing (phase modulation \(\propto \exp({-iku_{0}t})\)), the macroscopic density perturbations generated by this term decay according to a power law \(\sim 1/t\) due to the fact that
faster and slower particles fly apart. The second term describes the polarization of the Maxwell-Boltzmann medium: the initial momentum pushes or attracts background particles, exciting collective macroscopic oscillations (Van Kampen waves). The factor \(\frac{A(\mathbf{x},u)}{u-u_{0}}\)
shows that the collective response is strongest for particles whose velocities are close to the velocity of the initial momentum, \(u \approx u_0\).

Particular attention should be paid to the influence of the cosmological force imbalance (the \(\Lambda \) term) in the void. The Green's function is parameterized by the spatial coordinate \(\mathbf{x}\) inside the Maxwell-Boltzmann kernel
\(A(\mathbf{x}, u)\) and the functions \(\varepsilon^\pm(u, \mathbf{x})\). In the deep void (at the periphery, \(\mathbf{x} \to \infty\)) attributed to cosmological repulsion, the background density decreases and the core tends to zero: \(A(\mathbf{x}, u) \to 0\)
and \(\varepsilon^{\pm }\rightarrow 1\). Substituting this into the Green's function, we see that the second (collective) term completely disappears: $G_{void}(u,u_{0},t,\mathbf{x})\to \delta (u-u_{0}) \exp({-iku_{0}t})$, meaning that within the void, collective gravitational interactions vanish completely.

The point momentum evolves as in a completely empty space: the particles simply fly apart ballistically,
without generating any density waves and without experiencing resistance from the medium. In the gravitational core, that is, at the center of the structures, \(\mathbf{x} \to 0\), the quantity \(A(\mathbf{x}, u)\) is large, and the second term dominates.
Transitioning to the complex plane according to Landau's rule, the contribution ${\exp({-ikut})}/{\varepsilon^{+}}$ is transformed into discrete growing or decaying Jeans modes. A point pulse instantly induces a powerful collective capture of background particles, leading to an exponential growth of density perturbations (\(\sim \exp({\gamma_L t})\)) and initiating the process of local gravitational collapse.

\section{Stability of voids in the Local Universe}

To obtain an exact dispersion relation in a system with an external repulsive potential, the standard Fourier transform with respect to the spatial variable
is not directly applicable, since the unperturbed particle trajectories diverge exponentially, $d^2{\bf x}/dt^2=c^2\Lambda {\bf x}/3$. It is appropriate to use the method of integration over the unperturbed (by attraction) trajectories of particles ($dX/d\tau=V(\tau)$, $dV/d\tau=(c^2\Lambda/3) X(\tau)$, where $\tau = t-t_0$),
located at a given time $t$ at a point in phase space $({\bf x},{\bf v})$
backward in time to the moment $t_0<t$ (without a self-consistent field) via
\begin{equation}
X(\tau):\:\:{\bf x}(t_0)={\bf x}\cdot {\rm{cosh}}\big( \sqrt{c^2\Lambda/3} (t-t_0) \big) -
\end{equation}
$$
\frac{{\bf v}}{\sqrt{{c^2}{\Lambda}/3}}\cdot {\rm{sinh}}\big( \sqrt{c^2\Lambda/3} (t-t_0) \big),
\label{7}
$$
$$
V(\tau):\:\:{\bf v}(t_0)={\bf v}\cdot {\rm{cosh}}\big( \sqrt{c^2\Lambda/3} (t-t_0) \big) -
$$
$$
{{\bf x}}{\sqrt{c^2\Lambda/3}}\cdot {\rm{sinh}}\big( \sqrt{c^2\Lambda/3} (t-t_0) \big).
$$\noindent
Integrating the linearized Vlasov equation,
\begin{equation}
\frac{\partial f}{\partial t}+{\bf v}\nabla_{\bf x}{f}+(c^2\Lambda/3)\nabla_{\bf v}f=\nabla_{\bf x}\phi\cdot\nabla_{\bf v}F_0,
\label{8}
\end{equation}
\noindent
along the unperturbed characteristics, ignoring the influence of the homogeneous background $F_0$, from $\tau=0$ to $\tau=t$) yields, in accordance with Duhamel's principle,
$$
f({\bf x},{\bf v},t)=f(X(0),V(0),t_0=0)+
$$
$$
\int^t_0 \nabla_{\bf x}\phi \big( X(t_0), t_0 \big)\cdot \big(\partial F_0/\partial {\bf v\big)}\big|_{V(t_0)}dt_0.
$$

To obtain the dispersion relation, we seek spatio-temporal perturbations in the form of plane waves. However, due to the exponential expansion of the background, the wave vector,
\(\mathbf{k}\), must depend on time: ${\bf k}(t) = {\bf k}^\flat \exp({-\sqrt{c^2\Lambda/3} t})$. The general form of the potential is then
\begin{equation}
\phi({\bf x},t)=\phi^\flat \exp\bigg( i{\bf k}(t)\cdot {\bf x}-i\int \omega(t)\:dt \bigg).
\label{10}
\end{equation}
In the locally homogeneous approximation for short waves when the scale of the perturbation is much smaller than the scale of the background inhomogeneity, we can set $|{\bf{k}}| \approx {\rm{const}}$ and seek solutions in the form of normal van Kampen modes, as in classical plasma theory:
$f({\bf x},{\bf v},t)=f^\flat_k({\bf v})\exp(i{\bf k}\cdot{\bf x}-i\omega t)$ and
$\phi ({\bf x},t)=\phi^\flat_k \exp(i{\bf k}\cdot{\bf x}-i\omega t)$ (we assume $Im\:\omega >0$). In this case, on the characteristic we have
$\partial \phi/\partial {\bf x}=i{\bf k}\phi^\flat_k \exp\big(i{\bf k}{\bf x}|_{{\bf x} \to X(\tau)} - i\omega \tau \big)$. The expression for the disturbance of the distribution function takes the form
\begin{equation}
f({\bf x},{\bf v},t)= i{\bf k}\phi^\flat_k \cdot
\int^t_0 \exp\big( i {\bf k}{\bf x} \: {\rm{cosh}}(\sqrt{c^2\Lambda/3}t_0) -
\label{11}
\end{equation}
$$
-(i{\bf k}{\bf v}/\sqrt{c^2\Lambda/3}){\rm{sinh}}(\sqrt{c^2\Lambda/3}t_0)
-i\omega (t-t_0) \big)\frac{\partial F_0}{\partial {\bf v}}\big|_{{\bf v}\to V(\tau)}dt_0.
$$
Substituting the last expression into the Poisson equation for the self-consistent background, $-k^2\phi=4\pi \gamma \int f\:d{\bf v}$, yields the dispersion relation in the limit of constant eigen-vectors; in general case of background expansion, as said earlier, one should rely on formula (\ref{10}) for the potential. Indeed,
assuming $|{\bf x}|\to 0$ (i.e., the local limit), neglecting the coordinate dependence of the phase, we have
$$
f_k^\flat ({\bf v})=i\phi^\flat_k
\int^\infty_0 \big( {\bf k}\cdot \partial F_0 ({\bf v}_0)/\partial {\bf v}_0 \big)\exp\big( i\omega (t -t_0)-
$$
$$
i({\bf k}\cdot{\bf v}/\omega){\rm{sinh}}(\omega(t-t_0)) \big)dt_0.
$$
Substituting this limit into the Poisson equation $-k^2\phi_k^\flat = 4\pi \gamma \int f_k^\flat ({\bf v})d{\bf v}$ gives
\begin{equation}
1-4\pi i \gamma/k^2 \int_{{\mathbb R}^3} \int_0^\infty \big( {\bf k}\cdot \partial F_0 ({\bf v}_0)/\partial {\bf v}_0 \big)
\exp\big( i\omega \tau -
\end{equation}
$$
 i({\bf k}\cdot{\bf v}/\sqrt{c^2\Lambda/3}){\rm{sinh}}(\sqrt{c^2\Lambda/3}\tau) \big)
d\tau d{\bf v} =0,
\label{12}
$$
\noindent
where $\tau=0$ means $t_0=t$, $\tau\to \infty$ means $t_0\to -\infty$. Thus,
the effective permittivity of the medium \(\varepsilon(\omega, {\bf k})|_{|{\bf x}|\to 0}\) yields
\begin{equation}
\varepsilon(\omega, {\bf k})|_{|{\bf x}|\to 0} =
\label{13}
\end{equation}
$$
=1-4\pi i \gamma/k^2 \int_{{\mathbb R}^3} d{\bf v} \int_0^\infty d\tau \big( i{\bf k}\cdot \nabla_{{\bf v}_0} F_0 ({\bf v}_0) \big)
$$
$$
\exp\big( i\omega \tau - i({\bf k}\cdot{\bf v}/\sqrt{c^2\Lambda/3}){\rm{sinh}}(\sqrt{c^2\Lambda/3}\tau) \big).
$$
The velocity vector, \({\bf v}_0\), under the integral also depends on \(\tau\) according to the
characteristic formula, \({\bf v}_0 = {\bf v} \cosh(\sqrt{c^2\Lambda/3} \tau)\) (for \( |{\bf x}| \to 0\)). The influence of the repulsive term ($\sqrt{c^2\Lambda/3} >0$) fatally changes the behavior of the particle system. The presence of the term \(\sinh(\sqrt{c^2\Lambda/3} \tau)\) in the exponential radically changes the structure of the integral. Since $sinh (\sqrt{c^2 \Lambda/3}) \sim  \exp(\sqrt{c^2 \Lambda/3}/2)$ at large values of $\tau$, the phase oscillates with an exponentially increasing frequency. This leads to strong phase mixing (i.e., modified Landau damping). Ordinary undamped waves cannot exist in the system. The scale of the Jeans instability is restructured: for exponential growth of the perturbation, the gravitational instability increment
\(\Gamma = \text{Im}(\omega)\) must exceed the repulsion velocity, \(\sqrt{c^2\Lambda/3}\). Otherwise, the external field (i.e.,
the influence of the $\Lambda$ term) has time to prevent the formation of a gravitational blob due to rapid spatial dissipation.

Assuming that the equilibrium function, $F_0 ({\bf v})$, is set to be a Maxwellian via $F_0 \to F_M ({\bf v})=n_0 (2\pi v_T^2)^{-3/2}\exp(-v^2/v_T^2)$, we find that $n_0$ is the unperturbed particle density. In the approximation under consideration, $|{\bf x}|\to 0$ the Maxwellian permittivity takes the form
\begin{equation}
\varepsilon^M(\omega, {\bf k})|_{|{\bf x}|\to 0} =
1-\omega_J^2/(2\sqrt{c^2\Lambda/3})\cdot
\end{equation}
$$
\int_0^\infty {\rm{sinh(2\sqrt{c^2\Lambda/3}\tau)}}, 
$$
$$
\exp\big( i\omega\tau - \frac{k^2 v_T^2}{2c^2\Lambda/3} {\rm{sinh}}^2(\sqrt{c^2\Lambda/3}\tau) \big) d\tau,
\label{14}
$$
\noindent
where $\omega_J^2=4\pi \gamma n_0$ is the classical Jeans frequency. For $c^2\Lambda/3\to 0$; namely, in a self-gravitating medium without repulsion, we obtained an expression for the permittivity through the Kramp plasma dispersion function,
$$
\varepsilon^M(\omega, {\bf k})|_{|{\bf x}|,c^2\Lambda/3\to 0}\to 1- \omega_J^2/(k^2 v_T^2),
$$
$$
\big( 1+(\omega/k v_T)Z\big( \omega/ (\sqrt{2}kv_T)\big) \big).
$$\noindent
In Eq. (\ref{14}), for large values of $\tau$, we obtained ${\rm{sinh}}^2(\sqrt{c^2\Lambda/3}\tau)\to \exp(\sqrt{c^2\Lambda/3}\tau)$, then the integral will converge for any complex frequencies, $\omega$, and the Landau poles disappear. Using the Maxwellian distribution as an equilibrium solution leads to a remarkable conclusion regarding the existence of Jeans instability in the particle system whose dynamics is evolving both via gravity and repulsion. Here, this instability
is responsible for the exponential growth of perturbations ($Im(\omega)>0$). The instability boundary for the condition is $\omega\to 0$ and, in this limit, we transformed the expression (\ref{14}): ${{\lim}}_{\omega\to 0}\varepsilon(\omega,{\bf k})\to 1-\omega_J^2/(k^2 v_T^2)=0$.
Consequently, the value of the critical Jeans wave vector is $k_J=\omega_J/v_T=\sqrt{4\pi \gamma n_0}/v_T$,
while the critical Jeans wave is $\lambda_J=2\pi/k_J=v_T/\sqrt{\pi/(n_0\gamma)}$, identical to the critical scale in the absence of the influence of the $\Lambda$ term. However, external repulsion radically changes the dynamics at the instability scale (\(\lambda > \lambda_J\)); namely, for the perturbation to grow, the Jeans instability growth rate, \(\Gamma = \text{Im}(\omega)\), must be strictly greater than the velocity of external expansion,
\(\sqrt{c^2\Lambda/3} \). If \(\Gamma < \sqrt{c^2\Lambda/3}\), then the matter expands faster than the gravity can form a bound object.
An explicit criterion for perturbation growth (taking the growth rate into account) takes the form of a modified inequality: $\sqrt{\omega_J^2-k^2 v_T^2}>\sqrt{c^2\Lambda/3}$. The particle system, for the Maxwellian ``equilibrium'' function, becomes completely stable on all wave vectors. Specifically, the gravitational collapse is impossible at any scale, if the cosmological repulsive force exceeds the natural Jeans frequency, $\sqrt{c^2\Lambda/3}\ge \omega_J$.

As noted earlier in this work, to obtain exact expressions for the eigenstates of a system described by the linearized Vlasov--Poisson equation, instead of van Kampen modes, it is necessary to use the modes of the total eigenvalues, $f({\bf x},{\bf v},t)=f_\sharp ({\bf x},{\bf v})\exp(-i{\omega}t)$, $\phi({\bf x},t)=\phi_\sharp ({\bf x},{\bf v})\exp(-i\omega t)$.
Near the center of the system or in the approximation where the gravitational background \(\Phi_{0}\) is compensated for and the external structure dominates, the effective background potential
is quadratic: \(U_{\text{eff}}(\mathbf{x}) \approx \frac{1}{2} \Upsilon^2 \mathbf{x}^2\). Then, the density profile is a Gaussian. This is expressed as
$$n_0(\mathbf{x}) = n_0(0) \exp\big({-\Upsilon^2 \mathbf{x}^2 / 2v_T^2}\big),$$
$${\Upsilon}^2= {c^2 \Lambda /3} -\frac{1}{3}{\omega_J}^2({\bf x}).$$
We substituted this into the linearized Vlasov equation. Integrating over the velocities to obtain the density perturbation \(\rho_f = m \int f d^3v\), in the hydrodynamic (or macroscopic collisionless) approximation for quasi-coplanar or 1D modes along the chosen abscissa axis, we arrive at a second-order differential equation.

As established earlier, integrating the linearized Vlasov equation along unperturbed
trajectories backward in time (\(\tau = t - t_0\)) yields a solution for the perturbation,
\(f\). We note that for simplicity, we restricted our computation to a 1D geometry, with $x_0=x(t_0)$, $v_0=v(t_0)$), giving
$$f(x, v, t) = \int_{0}^{\infty} d\tau \, {\partial \phi_\sharp(x_0, t-\tau)}/{\partial{x_0}} \cdot {\partial F_0}/{\partial v(t-\tau)}.$$
We sought spectral modes of the form $\phi_\sharp(x, t) = \phi(x) \exp({-i\omega t})$ and similarly for \(f_\sharp\). Then \(\phi_\sharp(x_0, t-\tau) = \phi(x_0) \exp({-i\omega t}) \exp({i\omega \tau})\). The macroscopic density perturbation,
\(\rho_f(x)\), is found by integrating over velocities,
$$\rho_f(x) = m \int_{-\infty}^{\infty} f_\sharp(x, v)dv,
= m \int_{-\infty}^{\infty} dv,
$$
$$
\int_{0}^{\infty} d\tau \exp({i\omega \tau}) ({d \phi(x_0)}/{dx_0}) ({\partial F_0}/{\partial{v_0}}).
$$
For a spatially inhomogeneous Maxwellian, the derivative is \({\partial F_0}/{\partial v_0} = -\frac{v_0}{v_T^2} F_0(x_0, v_0)\).
Since the total energy of a particle along an unperturbed trajectory is conserved, the value of the background distribution function is constant: \(F_0(x_0, v_0) = F_0(x, v)\). We get\begin{equation}
\rho_f(x)=-(m/v^2_T)\int^\infty_{-\infty}dv\: F_0(x,v),
\end{equation}
$$
\int^\infty_0 d\tau \exp(i\omega\tau) v(\tau) \big(d\phi(x(\tau))/dx(\tau) \big),
\label{15}
$$
\noindent
where $x(\tau)=x\:{\rm{cosh}}(\Upsilon \tau) - (v/\Upsilon) {\rm{sinh}}(\Upsilon\tau)$, $v(\tau)= v\:{\rm{cosh}}(\Upsilon \tau) - \Upsilon x\: {\rm{sinh}}\:(\Upsilon\tau)$ ---
local trajectories of backward motion in time, which are expressed in terms of the current coordinates $(x,v)$. Expression (20) is nonlocal, as the value of $\rho_f (x)$ at a given point depends on the values of the potential perturbation $\phi$ at all points $x(\tau)$.
This relation can be reduced to an ordinary differential equation using an expansion in a small parameter. Thus, we expanded $x(\tau)$, $v(\tau)$ in the neighborhood of the point ${\tau =0}$, as well as the expression $d\phi (x(\tau))/dx(\tau)$ in the neighborhood of the current point $x$. Substituting the resulting expressions, to an accuracy of $\sim \tau^2$, into the right-hand side of the formula (\ref{15}), and simultaneously representing the density $\rho_f(x)$ from its left-hand side through the even modes of the Maxwellian (the local density $n_0({\bf x})=\int F_0d{\bf v}$ and the thermal velocity $\int v^2 F_0d{\bf v}=n_0({\bf x})v^2_T$), we have
$$
\rho_f ({\bf x}) =\frac{m}{\omega^2}\frac{d}{dx}\big( n_0 ({\bf x}) \frac{d\phi}{d{\bf x}} \big).
$$
\noindent
This enables us to rewrite the Poisson equation ${d^2\phi}/{dx^2}=4\pi \gamma \rho_f$ as
\begin{equation}
v^2_T \frac{d^2\phi}{dx^2} + \big( \omega^2 - (\omega_J^2 (x) - c^2\Lambda/3) \big)\phi=0,
\end{equation}
$$
\omega_J^2 (x) =4\pi \gamma m n_0 (x)=\omega_{J0}^2 \exp\big(-\Upsilon^2 x^2/(2v_T^2)\big).
\label{16}
$$
\noindent
It is assumed that the background density gradient is $(dn_0/dx)_{loc}\to 0$. Then, we can expand the density profile and, hence, the local Jeans frequency, near the center ($x \to 0$) in a Taylor series via
\(\omega_{J}^{2}(x)\approx \omega_{J0}^{2}\big(1-\frac{\Upsilon^{2}x^{2}}{2v_{T}^{2}}\big)\). So the equation takes the form
\begin{equation}
\frac{d^2\phi}{dx^2} + \big( v_T^{-2} ( \omega^2 - \omega_{J0}^2 (x) - c^2\Lambda/3) + v_T^{-4} \omega_{J0}^2\Upsilon^2 x^2/2 \big)\phi=0.
\label{17}
\end{equation}
\noindent
After a series of transformations, it turns out that we have the Weber equation for the potential: $\phi_{\xi\xi}''+(\mu+\xi^2/4)\phi=0$, $\mu=(c^2\Lambda/3-\omega^2_{J0}+\omega^2_n)/\sqrt{2\omega_{J0}^2\Upsilon^2}$, ${\xi} = {\beta} x$, ${\beta} v_T = (2 {\omega}_{J0}^2 \Upsilon^2)^{1/4}$.
The solutions of this equation are the parabolic cylinder functions (Weber functions) \(D_{\nu}(z)\). Since the sign in front of \(\xi^{2}\) is positive
(\(+\frac{1}{4}\)), this corresponds to the problem of oscillations in an effective potential well for wave disturbances. For the solution to be physical, the potential \(\phi\) must decay at infinity \(\xi \to \infty\), since the disturbances are localized inside the cloud, and quantization condition is imposed on the eigenvalues \(\mu \);
the standard condition of reality and mode localization requires that $\mu = n+1/2$, $n=0,1,2,...$. Therefore,
the discrete spectrum of the global eigen-frequencies of the particle system is
$$
(c^2\Lambda/3-\omega^2_{J0}+\omega^2_n)/\sqrt{2\omega^2_{J0}\Upsilon^2}=n+1/2.
$$
Unlike the continuous spectrum (plane waves in a homogeneous gas), the inhomogeneous Maxwellian generates a discrete set of macroscopic global modes (i.e., spheroidal or chain oscillations of a gravitating gas ball). We can consider a precise criterion for global instability: the mode with \(n=0\) (the fundamental mode) becomes unstable first when \(c^2\Lambda/3\) begins to dominate. The global stability of the entire system is determined by the sign of \(\omega_{n}^{2}\). If the repulsive force due to the cosmological term is strong enough, even the central core (\(\omega_{J0}\)) is unable to hold the system together, and the entire structure enters a global decay mode.

To analyze a physically correct solution describing spatially localized perturbations, we ought to consider the behavior of the potential at large distances from the center of the system (\(\xi \to \infty\)), where the repulsion dominates. Using the WKB method yields the following asymptotics of the potential,
$$
\lim_{\xi \to \infty}\phi (\xi)= C\xi^{-1/2}\exp\big(\pm i (\xi^2/4 +\mu\:\ln(\xi) ) \big).
$$
\noindent
Returning to the dimensional coordinate $x=\xi/\beta$, where $\beta \equiv (2\omega_{J0}^2\Upsilon^2)^{1/4}/v_T$, we can draw a number of remarkable physical conclusions: the potential $\phi (x)$ decreases at large distances according to a power law of $\sim x^{-1/2}$ and, at the same time,
the wave phase increases proportionally to the square of the coordinate \(\propto x^{2}\).
The latter can be interpreted as follows: density perturbations in the repulsive potential form "runaway" waves and particles, escaping to the periphery, are accelerated by the external field, causing the wavelength of the perturbations to rapidly decrease (\(\lambda \sim 1/x\)). In addition, the density of the wave perturbation is smeared across space. The potential decays purely geometrically due to the expansion of matter and there are no correlations with the Landau decay involved here.

In a cosmology without dark energy, at $c^2 \Lambda /3 =0$, any perturbations with a scale greater than the Jeans scale (\(k < k_J\)) grow exponentially or in a way resembling a power law over time, assumed to form galaxies and their clusters. The introduction of repulsion ($\Lambda > 0$) radically changes the picture on large spatial scales (small \(k\)). Our criterion $\omega_J^2 < c^2 \Lambda /3 + \dots$  shows that in regions of reduced matter density, where \(\rho_m\) is small,
the repulsion of dark energy completely suppresses the gravitational instability. In such cases, perturbations cannot grow. This explains why cosmic voids remain stable and do not collapse, continuing to expand exponentially. Furthermore, when the Universe transitions to the radiation-dominated or dark energy-dominated stage, the growth of large-scale density fluctuations stops, as frozen. The system loses the ability to generate new bound gravitational objects.

The discrete spectrum of global eigenmodes, \(\omega_{n}^{2}\), that we derived, together with the spatial asymptotics of the Weber functions, describes the dynamics of isolated cosmic halos (e.g., galaxy clusters) embedded in the accelerating background of the Universe. The slow power-law decay of the potential, \(\propto 1/\sqrt{x}\), and the quadratic growth of the wave phase at infinity characterize the gradual escape of matter from the cluster. 

From Eq. (1) follows the existence of a critical radius, 
$$
r_{crit}= \big(3\gamma m/\Lambda c^2\big)^{1/3},
$$
at which the gravitational attraction of the system is exactly balanced by the repulsive effect of the cosmological constant. Within this radius, matter remains in a quasi-stationary Maxwellian equilibrium described by the discrete modes, \(n\). Beyond $r_{crit}$, particles become trapped by the cosmological Weber barrier and are carried away by the global Hubble expansion of the Universe, forming a regime of runaway asymptotic waves.

\section{Conclusions}

Within the framework of the Vlasov kinetic approach, we are able to analyze the structure of the configurations in the Local Universe that are not only governed by gravitational interaction, but also given by the terms of Eq. (1) with the repulsion term of the cosmological constant or of the dark energy on that scales. Our analysis has shown that the conditions inside the void correspond to the regime of $\Lambda$-domination,
$$
 \frac{\Lambda c^2}{3} > 4\pi \gamma \rho_{void},
$$
leading to the following principal conclusions:

1. Stability of the void structure: As shown from the Fredholm equation of the third kind, the Jeans increment in this region becomes imaginary, while the Hubble friction together with enhanced Landau damping completely suppresses any discrete collapse modes. Consequently, random local density perturbations inside the void cannot grow to involve new galaxies to the walls: they are collisionlessly smeared out by thermal motion and dispersed by cosmological repulsion.

2. The crucial role of the cosmological constant repulsion (local dark energy): Inside the void, the $\Lambda$-repulsion exceeds the attractive force of the residual matter. As a result, the void behaves not merely as an “empty” region, but as a dynamical object with an effectively negative gravitational mass. It actively repels the surrounding matter, thereby accelerating the migration of matter toward the void boundaries. 

Thus, we conclude that in the present epoch, the late Universe, the formation of new voids has practically ceased, while the existing voids have entered a stage of saturation characterized by stable and more pronounced walls. We also note that recent detailed analyses of observational data indicate a predominance of voids with increasingly pronounced walls at lower redshifts \citep{Tav}. The latter analysis is remarkable both with respect to the properties of particular galaxy clusters and their environments \citep{As1,As2} and for testing the predictions of the $\Lambda CDM$ model. 

\begin{acknowledgements}
We are thankful to the referee for useful comments.
\end{acknowledgements}

\end{document}